\documentclass[fleqn,usenatbib,useAMS]{Papers}
\usepackage{newtxtext,newtxmath}
\usepackage[T1]{fontenc}
\usepackage{subfloat}
\usepackage{makecell}
\usepackage[authoryear]{natbib}
\usepackage{graphicx}	
\usepackage{amsmath}	

\title[Halo mass functions from maximum entropy]{Halo mass functions from maximum entropy distributions in collisionless dark matter flow}

\author[Z. Xu]{
Zhijie (Jay) Xu,$^{1}$\thanks{E-mail: zhijie.xu@pnnl.gov; zhijiexu@hotmail.com}
\\
$^{1}$Physical and Computational Sciences Directorate, Pacific Northwest National Laboratory; Richland, WA 99352, USA\\
}

\date{Accepted XXX. Received YYY; in original form ZZZ}

\pubyear{2022}

\begin{document}
\label{firstpage}
\pagerange{\pageref{firstpage}--\pageref{lastpage}}
\maketitle

\begin{abstract}
The halo-mediated inverse mass cascade is a key feature of the intermediate statistically steady state for self-gravitating collisionless dark matter flow (SG-CFD). A broad spectrum of halos and halo groups are necessary to form from inverse mass cascade for long-range interaction system to maximize its entropy. The limiting velocity ($\textbf X$), speed ($\textbf Z$), and energy ($\textbf E$) distributions of collisionless particles can be obtained analytically from a maximum entropy principle. Halo mass function, the distribution of total mass in halos, is a fundamental quantity for structure formation and evolution. Instead of basing mass functions on simplified spherical/elliptical collapse models, it is possible to reformulate mass function as an intrinsic distribution to maximize system entropy during the everlasting statistically steady state. Starting from halo-based description of non-equilibrium dark matter flow, distributions of particle virial dispersion ($\textbf H$), square of particle velocity ($\textbf P$), and number of halos ($\textbf J$) are proposed. Their statistical properties and connections with velocity distribution ($\textbf X$) are well studied and established. With $\textbf H$ being essentially the halo mass function, two limiting cases of $\textbf H$ distribution are analyzed for large halos ($\textbf H_\infty$) and small halos ($\textbf H_s$), respectively. For large halos, $\textbf H_\infty$ is shown to also be a maximum entropy distribution. For small halos, $\textbf H_s$ approximates the $\textbf P$ distribution and recovers the Press-Schechter mass function. The full solution of $\textbf H$ distribution is determined by the velocity distribution ($\textbf X$) that maximizes system entropy and the exact model of halo velocity dispersion.
\end{abstract}

\begin{keywords}
\vspace*{-10pt}
Dark matter; N-body simulations; Halo mass function; Maximum entropy
\end{keywords}

\begingroup
\let\clearpage\relax
\tableofcontents
\endgroup

\section{Introduction}
\label{sec:1}
The large-scale nonlinear structure formation and evolution is essentially a self-gravitating collisionless fluid dynamics problem (SG-CFD) for dark matter flow. The gravitational instability leads to the self-organizing of collisionless dark matter particles into structures on different scales. The formation of halo structure is a major manifestation of the nonlinear gravitational collapse \citep{Neyman:1952-A-Theory-of-the-Spatial-Distri,Cooray:2002-Halo-models-of-large-scale-str}. For long-range interaction system, it is necessary to form halos and halo groups of different size to maximize system entropy \citep{Xu:2021-The-maximum-entropy-distributi}. The distribution of total mass in all halos, i.e. halo mass function, is one of the most fundamental quantities for analytical and/or semi-analytical modeling of structure formation and evolution. 

The first landmark of mass function might be the Press-Schechter (PS) formalism \citep{Press:1974-Formation-of-Galaxies-and-Clus,Bond:1991-Excursion-Set-Mass-Functions-f} that allows one to predict the shape and the evolution of the halo mass function. The PS model assumes objects collapse spherically and grow hierarchically from small, initially Gaussian density fluctuations. Objects will collapse at some mass scale once the smoothed linear density contrast on that scale exceeds a threshold value $\delta _{c} $. This value can be analytically derived by examining the nonlinear collapse of a spherical top-hat over-density \citep{Tomita:1969-Formation-of-Gravitationally-B,Gunn:1972-Infall-of-Matter-into-Clusters} or the two-body collapse problem, an elementary step in inverse mass cascade \citep{Xu:2021-A-non-radial-two-body-collapse,Xu:2021-Inverse-mass-cascade-mass-function}. 

The spherical collapse model, a simple but very powerful analytical tool for the non-linear evolution of structures, predicts the value of  $\delta _{c} $ on the order of unity and independent of the collapsed object size or mass. The exact same threshold value $\delta _{c} $ can be also obtained by a recently proposed two-body collapse model (TBCM) that mimic the harmonic oscillator model for dynamics \citep{Xu:2021-A-non-radial-two-body-collapse}. The threshold values $\delta _{c} $ corresponds to the density of growing halos with extremely fast mass accretion. Such halos should have an isothermal density profile \citep{Xu:2021-A-non-radial-two-body-collapse}. In practice, halos have finite mass accretion rate and density profile cannot be isothermal due to halo deformation along radial direction. The effect of mass cascade on halo density is also formulated \citep{Xu:2021-Inverse-mass-cascade-halo-density}, where a random walk of collisionless particles in a dynamically varying halo is presented.

When a normalized variable $\nu ={\delta _{c}^{2} /\sigma _{\delta }^{2} \left(m_{h} \right)} $ is introduced, the PS mass function can be written compactly as
\begin{equation} 
\label{ZEqnNum333244} 
f_{PS} \left(\nu \right)=\frac{1}{\sqrt{2\pi } \sqrt{\nu } } e^{-{\nu /2} } ,         
\end{equation} 
where $\sigma _{\delta }^{2} \left(m_{h} \right)$ is the density fluctuation when density field is smoothed at halo mass scale $m_{h} $.  The multiplicity mass function is expressed as $f_{PS}^{m} \left(\nu \right)=2f_{PS}^{} \left(\nu \right)\nu $. Bond et al. \citep{Bond:1991-Excursion-Set-Mass-Functions-f} provided an alternative derivation of the PS model using an excursion set approach (Extended PS or EPS model). The excursion set formalism puts the theory on a firmer footing by removing the fudge factor introduced in the original PS model. Two assumptions were made in EPS: i) the threshold overdensity was computed using the spherical collapse model; ii) the linear overdensity at a given location in space is assumed to vary with a smoothing scale as a random walk process when a sharp \textit{k}-space filter is used for the smoothing. 

However, when compared to numerical simulations, it was found that both PS and EPS models do not exactly match the results of \textit{N}-body simulations \citep{Jenkins:2001-The-mass-function-of-dark-matt}. While agree with the simulation data at current epoch reasonably well, both models overpredict the number of low-mass halos and underpredict the number of massive halos. There are also significant errors at high redshifts \citep{Springel:2005-Simulations-of-the-formation--}. Further improvement was achieved by relaxing the first assumption in EPS model and computing the density threshold for ellipsoidal collapse \citep{Sheth:2001-Ellipsoidal-collapse-and-an-im,Sheth:1999-Large-scale-bias-and-the-peak-}. In contrast to the spherical collapse where the threshold $\delta _{c} $ is independent of the mass scale, the ellipsoidal collapse model gives a mass-dependent overdensity threshold (a moving barrier). This modification considerably complicates the original model derivation but was shown to yield a better agreement with simulations. The modified PS model (ST model) can be compactly written as:
\begin{equation} 
\label{ZEqnNum971057} 
f_{ST} \left(\nu \right)=A\sqrt{\frac{2q}{\pi } } \left(1+\frac{1}{\left(q\nu \right)^{p} } \right)\frac{1}{2\sqrt{\nu } } e^{-{q\nu /2} } ,       
\end{equation} 
where the normalization condition requires: 
\begin{equation} 
\label{eq:3} 
A=\frac{\sqrt{\pi } }{\Gamma \left({1/2} \right)+2^{-p} \Gamma \left({1/2} -p\right)} .         
\end{equation} 
The best fitted parameters from simulation is $A=0.3222$, $q=0.75$, and $p=0.3$ \citep{Sheth:2002-An-excursion-set-model-of-hier}. It is obvious that with $A=0.5$, $q=1.0$, and $p=0$, the ST mass function reduces to the original PS function in Eq. \eqref{ZEqnNum333244}. Both ST and PS model satisfy the normalization condition $\int _{0}^{\infty }f\left(\nu \right) d\nu =1$ that requires all mass belongs to halos.

Since the halo formation and evolution is an extremely complicated nonlinear process, direct numerical simulations become crucial to drive the development of theory. Many forms of empirical mass functions were also proposed by fitting to the high-resolution simulation data \citep{Warren:2006-Precision-determination-of-the,Reed:2007-The-halo-mass-function-from-th}. For example, a universal mass function covers a wide range of simulations with different cosmologies and redshifts \citep{Jenkins:2001-The-mass-function-of-dark-matt},
\begin{equation} 
\label{eq:4} 
f_{JK} \left(\nu \right)=\frac{0.315}{2\nu } \exp [-\left|\ln \left({\sqrt{v} /\delta _{c} } \right)+0.61\right|^{3.8} ] ,       
\end{equation} 
where $\delta _{c} =1.6865$ at $z=0$. It should be noted that these empirical mass functions might not satisfy the normalization constraint and can be difficult to extrapolate beyond the range of fit.

Recently, a new form of (double-$\lambda$) mass function is proposed based on the inverse mass cascade theory for dark matter flow, a type of SG-CFD \citep{Xu:2021-Inverse-mass-cascade-mass-function}. The mass and energy cascades \citep{Xu:2021-Inverse-mass-cascade-mass-function,Xu:2021-Inverse-and-direct-cascade-of-} is a fundamental feature to understand the evolution of halo energy and momentum \citep{Xu:2022-The-mean-flow--velocity-disper,Xu:2022-The-evolution-of-energy--momen} and develop the statistical theory for dark matter flow \citep{Xu:2022-The-statistical-theory-of-2nd,Xu:2022-The-statistical-theory-of-3rd,Xu:2022-Two-thirds-law-for-pairwise-ve}. In addition, it is also potentially relevant to the dark matter particle mass and properties \citep{Xu:2022-Postulating-dark-matter-partic}, MOND (modified Newtonian dynamics) theory \citep{Xu:2022-The-origin-of-MOND-acceleratio}, and baryonic-to-halo mass relation \citep{Xu:2022-The-baryonic-to-halo-mass-rela}.  

Mass cascade has two distinct regimes: 1) propagation range where mass is simply propagated by halos to larger scales for halos of mass $m_{h} <m_{h}^{*} $; 2) deposition range where mass is actively consumed to grow halos for halos of mass $m_{h} >m_{h}^{*} $, where $m_{h}^{*} $ is a characteristic mass scale. Entire mass cascade can be formulated by the random walk of halos in mass space, where halos migrate via merging with "single mergers". The waiting time $\tau _{g} $ of random walk (halo lifetime $\tau _{g} $) is dependent on a geometry parameter $\lambda $ as $\tau _{g} \sim m_{h}^{-\lambda } $ \citep{Xu:2021-Inverse-mass-cascade-mass-function}. In this theory, mass function can be analytically obtained without relying on any specific spherical /elliptical collapse models. Two different values of parameter $\lambda $ for two regimes of inverse mass cascade lead to a new simple mass function, the so-called double-$\lambdaup$ mass function \citep[see][Eq. (98)]{Xu:2021-Inverse-mass-cascade-mass-function}.
\begin{equation} 
\label{ZEqnNum982412} 
f_{D\lambda } \left(\nu \right)=\frac{\left(2\sqrt{\eta _{0} } \right)^{-q} }{\Gamma \left({q/2} \right)} \nu ^{{q/2} -1} \exp \left(-\frac{\nu }{4\eta _{0} } \right),       
\end{equation} 
with values of $\eta _{0} =0.76$ and $q=0.556$ for the best fit to simulation. 

Because of the simplicity, the PS-EPS-ST mass functions are still the only and the most popular analytic models for the formation, distribution, and evolution of halos. However, the theoretical basis of this approach is at best heuristic. First, the entire derivation requires a threshold overdensity that must be calculated based on a simplified (if not over simplified) collapse model (either spherical or ellipsoidal collapse models). Second, the linear density field is required to identify collapsed structures that is deeply in the non-linear regime. Finally, a specific smoothing filter (the sharp \textit{k}-space filter) is required for the random walk of a local overdensity. 

In principle, halo mass function should be an objective and intrinsic property of self-gravitating collisionless system involving long-range interactions, independent of the choice of collapse models or smoothing filters. A different view of halo mass functions seem to be necessary and enlightening. The mass function is a probability distribution due to the random walk of halos in mass space, a direct result of the inverse mass cascade \citep{Xu:2021-Inverse-mass-cascade-mass-function}. 

Since inverse mass cascade is necessary for self-gravitating system to generate a broad spectrum of halos and maximize system entropy \citep{Xu:2021-The-maximum-entropy-distributi}, mass function should also be a direct result of entropy maximization for non-equilibrium dark matter flow at statistically steady state. It is natural to ask what is the fundamental role of mass function from a statistical mechanics point of view. Instead of developing new mass functions, In this paper we focus on revealing the intrinsic connections of mass function with other maximum entropy distributions in dark matter flow. This represents a new point of view of halo mass function and its role in statistical mechanics, where mass function is proposed to be an intrinsic distribution to maximize the entropy of non-equilibrium system. 

\section{Limiting probability distributions in dark matter flow}
\label{sec:2}
\subsection{The problem settings}
\label{sec:2.1}
Consider a self-gravitating fluid consisting of \textit{N} collisionless particles interacting through a two-body power-law potential $V\left(r\right)$, i.e. $V_{} \left(r\right)\propto r^{n} $, where \textit{n} is an exponent of potential with $n=-1$ for the standard gravitational interaction. The spatial distribution of collisionless particles at statistically steady state is made up of distinct halos with a range of different sizes. A full statistical description of the entire system requires the knowledge of the distribution of halo size, the distribution of particles in individual halos, and the spatial clustering of halos. The distribution of the halo size (mass function) is the focus of current paper.

\begin{figure}
\includegraphics*[width=\columnwidth]{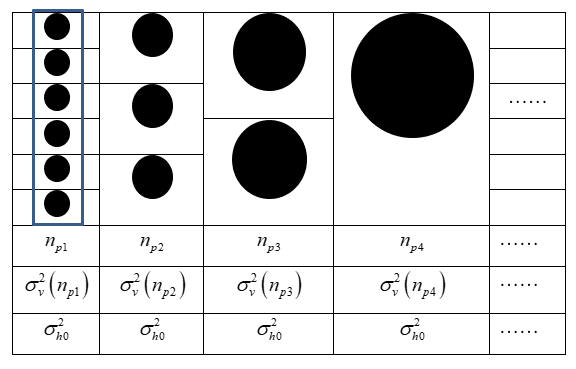}
\caption{The schematic plot of all halos of different size at statistically steady state. Halos are grouped and sorted according to the number of particles $n_{p} $ in the halo with increasing size from left to right. Every halo group is characterized by a halo virial dispersion $\sigma _{v}^{2} \left(n_{p} \right)$ and a halo velocity dispersion $\sigma _{h}^{2} $ that is relatively independent of halo size.}
\label{fig:1}
\end{figure}

The halo description of the statistically steady state (mostly studied for $n=-1$) is a natural result of entropy maximization for self-gravitating collisionless system with long-range interaction \citep{Xu:2021-The-maximum-entropy-distributi}. This description is presumably extended to the collisionless flow with exponent $n\ne -1$ but the interaction is still long-range. Figure \ref{fig:1} is a schematic plot of the halo picture by sorting the halos according to their sizes from smallest to largest halos. Each column is a group of halos of the same size. The statistics can be defined on three different levels: 1) individual halos; 2) group of halos of the same size (columns in Fig. \ref{fig:1}); and 3) global system including all particles from all halos. 

At the halo group level, the virial equilibrium is assumed. Halo group of size $n_{p}^{} $ has a halo virial dispersion $\sigma _{v}^{2} $ (the average temperature of all halos in the same group),
\begin{equation} 
\label{ZEqnNum386172} 
\sigma _{v}^{2} \propto \frac{Gm_{h} }{r_{h}^{-n} } \propto m_{h}^{{1+n/3} } =m_{h}^{{1/\beta } } \propto n_{p}^{{1+n/3} } =n_{p}^{{1/\beta } } ,       
\end{equation} 
where $m_{h} =n_{p} m_{p} \propto r_{h}^{3} $ is the mass of halos in the same group, $n_{p} $ is the number of collisionless particles in halo, $m_{p} $ is the mass of a single particle, $r_{h} $ is a characteristic (virial) size, and $\beta $ is the exponent between halo mass and halo virial dispersion $m_{h} \propto \left(\sigma _{v}^{2} \right)^{\beta}$. 

At the group level, particle velocity of all particles in the same halo group follows Maxwell-Boltzmann statistics due to the virial equilibrium \citep{Xu:2021-The-maximum-entropy-distributi}. Gaussian velocity distribution is expected for particles in the same group. The total particle velocity dispersion can be decomposed into:
\begin{equation} 
\label{ZEqnNum805703} 
\sigma _{}^{2} \left(m_{h} \right)=\sigma _{v}^{2} \left(m_{h} \right)+\sigma _{h}^{2} \left(m_{h} \right),         
\end{equation} 
with two separate contributions from the halo virial dispersion $\sigma _{v}^{2} $ and halo velocity dispersion $\sigma _{h}^{2} $, respectively. Here $\sigma _{h}^{2} $ is the one-dimensional halo velocity dispersion that is defined as the dispersion (variance) of halo velocity for all halos in the same group, i.e. the temperature of halo group due to the motion of halos \citep[see][Eqs. (2) and (3)]{Xu:2021-The-maximum-entropy-distributi}. In principle, both velocity dispersions can be functions of halo size $n_{p} $ or mass $m_{h} $. The two dispersions, $\sigma _{v}^{2} $ and $\sigma _{h}^{2} $, scale very differently with halo size. Here $\sigma _{v}^{2} \propto \left(m_{h} \right)^{{1/\beta}}$, while $\sigma _{h}^{2}$ is relatively independent of halo size (see Fig. \ref{fig:2}). 

At the global system level, the self-gravitating system can be characterized by the total number of collisionless particles \textit{N} and the one-dimensional velocity dispersion $\sigma _{0}^{2} $ of all particles (a measure of the total kinetic energy of entire system).

\subsection{Limiting probability distributions and their relations}
\label{sec:2.2}
For system described above, several limiting probability distributions can be identified:

\begin{enumerate}
\item \noindent $X(v)$: Distribution of one-dimensional particle velocity \textit{v};
\item \noindent $Z(v)$: Distribution of particle speed (the magnitude of the velocity vector);
\item \noindent $E(\varepsilon)$: Distribution of particle energy $\varepsilon $ including potential and kinetic energy;
\item \noindent $H(\sigma _{v}^{2})$: Distribution of particles with a given virial dispersion $\sigma _{v}^{2}$; Particle virial dispersion is the virial dispersion $\sigma _{v}^{2}$ of the halo group they belong to. 
\item \noindent $J(\sigma _{v}^{2})$: Distribution of number of halos with a given virial dispersion $\sigma _{v}^{2} $; Halo's virial dispersion is the virial dispersion $\sigma _{v}^{2} $ of the halo group they belong to.
\item \noindent $P(v^{2})$: Distribution of square of one-dimensional velocity \textit{v}; 
\end{enumerate}

First three (the \textit{X}, \textit{Z} and \textit{E}) distributions have been discussed in previous paper \citep{Xu:2021-The-maximum-entropy-distributi}, while the rest three (the \textit{H}, \textit{J}, and \textit{P}) distributions will be studied in this paper. Among these distributions, the relationship between the distributions \textit{X} and \textit{H} can be expressed as an integral transformation,
\begin{equation} 
\label{ZEqnNum517147} 
X\left(v\right)=\int _{0}^{\infty }\frac{1}{\sqrt{2\pi } \sigma } e^{-{v^{2} /2\sigma ^{2} } } H\left(\sigma _{v}^{2} \right)d\sigma _{v}^{2}  ,       
\end{equation} 
where the velocity distribution (\textit{X} distribution) is written as a weighted average of Gaussian distribution of particle velocity in different halo groups. Particle velocity dispersion $\sigma^{2} $ for a halo group is given by Eq. \eqref{ZEqnNum805703}. From Eq. \eqref{ZEqnNum517147}, second order moments of \textit{X} and \textit{Z} distributions are related by
\begin{equation} 
\label{ZEqnNum181268} 
\int _{-\infty }^{\infty }X\left(v\right)v^{2} dv =\int _{0}^{\infty }H\left(\sigma _{v}^{2} \right)\sigma ^{2} d\sigma _{v}^{2}  =\sigma _{0}^{2} .       
\end{equation} 
The mean halo virial dispersion and velocity dispersion for all halos in the system are (Eq. \eqref{ZEqnNum805703}),
\begin{equation} 
\label{ZEqnNum177519} 
\left\langle \sigma _{v}^{2} \right\rangle =\int _{0}^{\infty }H\left(\sigma _{v}^{2} \right)\sigma _{v}^{2} d\sigma _{v}^{2}  ,         
\end{equation} 
\begin{equation} 
\label{ZEqnNum435332} 
\left\langle \sigma _{h}^{2} \right\rangle \equiv \bar{\sigma }_{h}^{2} =\int _{0}^{\infty }H\left(\sigma _{v}^{2} \right)\sigma _{h}^{2} d\sigma _{v}^{2}  ,         
\end{equation} 
where $\bar{\sigma }_{h}^{2} $ is the mean halo velocity dispersion of entire system. It is easy to confirm that the mean velocity dispersion of all particles,
\begin{equation} 
\label{eq:12} 
\left\langle \sigma _{}^{2} \right\rangle =\left\langle \sigma _{v}^{2} \right\rangle +\left\langle \sigma _{h}^{2} \right\rangle =\sigma _{0}^{2} .         
\end{equation} 
Next, the relation between \textit{J} and \textit{H} distributions can be found from
\begin{equation}
\label{ZEqnNum350237} 
N_{h} \bar{N}H\left(\sigma _{v}^{2} \right)d\sigma _{v}^{2} =N_{h} J\left(\sigma _{v}^{2} \right)d\sigma _{v}^{2} n_{p} \left(\sigma _{v}^{2} \right),       
\end{equation} 
where both sides of Eq. \eqref{ZEqnNum350237} describe the number of particles in a halo group with a virial dispersion between $[\sigma _{v}^{2} ,\sigma _{v}^{2} +d\sigma _{v}^{2}]$. Here $N_{h} $ is the total number of halos in the system and $\bar{N}$ is the average number of particles per halo. In Eq. \eqref{ZEqnNum350237}, halo size $n_{p}(\sigma _{v}^{2})$ is a function of virial dispersion $\sigma _{v}^{2}$ and $H(\sigma _{v}^{2})d\sigma_{v}^{2}$ is the fraction of particles with a virial velocity dispersion between $[\sigma _{v}^{2} ,\sigma _{v}^{2} +d\sigma _{v}^{2} ]$. Here $J(\sigma_{v}^{2})d\sigma_{v}^{2} $ is the fraction of halos with a virial dispersion between $[\sigma _{v}^{2} ,\sigma _{v}^{2} +d\sigma _{v}^{2} ]$. From Eq. \eqref{ZEqnNum350237}, we have
\begin{equation} 
\label{ZEqnNum639447} 
H\left(\sigma _{v}^{2} \right)=J\left(\sigma _{v}^{2} \right){n_{p} \left(\sigma _{v}^{2} \right)/\bar{N}} ,         
\end{equation} 
and the average number of particles per halo is
\begin{equation} 
\label{eq:15} 
\bar{N}=\int _{0}^{\infty }J\left(\sigma _{v}^{2} \right)n_{p} \left(\sigma _{v}^{2} \right)d\sigma _{v}^{2}  .         
\end{equation} 

The \textit{P} distribution for the square of one-dimensional particle velocity ($v^{2} $) can be related to the \textit{H} distribution through an integral transformation (from Eq. \eqref{ZEqnNum517147}), 
\begin{equation}
\label{ZEqnNum757414} 
P\left(x=v^{2} \right)=\frac{X\left(\sqrt{x} \right)}{\sqrt{x} } =\int _{0}^{\infty }\underbrace{\frac{1}{\sqrt{2\pi x} \sigma } e^{-{x/2\sigma ^{2} } } }_{1}H\left(\sigma _{v}^{2} \right)d\sigma _{v}^{2}  ,     
\end{equation} 
where term 1 of Eq. \eqref{ZEqnNum757414} is a (one degree of freedom) Chi-square distribution for a halo group with a given virial dispersion $\sigma _{v}^{2} $ or total dispersion $\sigma ^{2} $. The Laplace transform of \textit{X} and \textit{P} distributions and the \textit{m}th order moments can all be related to the \textit{H} distribution via Eqs. \eqref{ZEqnNum517147} and \eqref{ZEqnNum757414},  
\begin{equation}
\int _{-\infty }^{\infty }X\left(v\right) e^{-vt} dv=\int _{0}^{\infty }H\left(\sigma _{v}^{2} \right)e^{{\sigma ^{2} t^{2} /2} } d\sigma _{v}^{2},      
\label{ZEqnNum803629}
\end{equation}
\begin{equation} 
\label{ZEqnNum887505} 
\int _{0}^{\infty }P\left(x\right) e^{-xt} dx=\int _{0}^{\infty }H\left(\sigma _{v}^{2} \right)\frac{1}{\sqrt{1+2\sigma ^{2} t} } d\sigma _{v}^{2}  ,       
\end{equation} 
\begin{equation} 
\label{ZEqnNum862024} 
\int _{-\infty }^{\infty }X\left(v\right) v^{m} dv=\frac{2^{{m/2} } }{\sqrt{\pi } } \Gamma \left(\frac{m+1}{2} \right)\int _{0}^{\infty }H\left(\sigma _{v}^{2} \right)\sigma ^{m} d\sigma _{v}^{2}  ,      
\end{equation} 
\begin{equation} 
\label{eq:20} 
\int _{0}^{\infty }P\left(x\right) x^{m} dz=\frac{2^{m} }{\sqrt{\pi } } \Gamma \left(\frac{1}{2} +m\right)\int _{0}^{\infty }H\left(\sigma _{v}^{2} \right)\sigma ^{2m} d\sigma _{v}^{2}  .      
\end{equation} 

The limiting distribution of one-dimensional velocity (\textit{X}) reads \citep[see][Eq. (32)]{Xu:2021-The-maximum-entropy-distributi}
\begin{equation} 
\label{ZEqnNum864132} 
X\left(v\right)=\frac{1}{2\alpha v_{0} } \frac{e^{-\sqrt{\alpha ^{2} +\left({v/v_{0} } \right)^{2} } } }{K_{1} \left(\alpha \right)} ,      
\end{equation} 
where $K_{y} \left(x\right)$ is a modified Bessel function of the second kind. The velocity $v_{0}$ is a typical scale of velocity and $\alpha $ is a shape parameter. The distribution $X\left(v\right)$ approaches a double-sided Laplace distribution with $\alpha \to 0$ and a Gaussian distribution for $\alpha \to \infty$. For an intermediate value of \textit{$\alpha$}, 

\begin{equation}
X\left(v\right)=\frac{e^{-\alpha } }{2\alpha v_{0} K_{1} \left(\alpha \right)} \exp \left(-\frac{v^{2} }{2\alpha v_{0}^{2} } \right) \quad \textrm{for} \quad v\ll v_{0}    
\label{ZEqnNum174426}
\end{equation}
\noindent and
\begin{equation}
X\left(v\right)=\frac{1}{2\alpha v_{0} K_{1} \left(\alpha \right)} \exp \left(-\frac{v}{v_{0} } \right) \quad \textrm{for} \quad v\gg v_{0}.    
\label{ZEqnNum367833}
\end{equation}

\noindent The \textit{X} distribution has a Gaussian core for small velocity $v$ (with a variance of $\alpha v_{0}^{2} $) and exponential wings for large velocity $v$. This feature is also observed from many large-scale \textit{N}-body simulations \citep{Cooray:2002-Halo-models-of-large-scale-str}. From Eq. \eqref{ZEqnNum181268}, we have the second moment of \textit{X} distribution 
\begin{equation} 
\label{eq:24} 
\sigma _{0}^{2} =\int _{-\infty }^{\infty }X\left(v\right)v^{2} dv =\alpha \frac{K_{2} \left(\alpha \right)}{K_{1} \left(\alpha \right)} v_{0}^{2} .        
\end{equation} 
The moment-generating function of \textit{X} distribution and the \textit{n}th order moments can be found as
\begin{equation} 
\label{ZEqnNum773894} 
MGF_{X} \left(t\right)=\int _{-\infty }^{\infty }X\left(v\right)e^{vt}  dv=\frac{K_{1} \left(\alpha \sqrt{1-\left(v_{0} t\right)^{2} } \right)}{K_{1} \left(\alpha \right)\sqrt{1-\left(v_{0} t\right)^{2} } } ,       
\end{equation} 
\begin{equation} 
\label{eq:26} 
\begin{split}
M_{X} \left(m\right)&=\int _{-\infty }^{\infty }X\left(v\right)v^{m}  dv\\
&=\frac{\left(2\alpha \right)^{{m/2} } \Gamma \left({\left(1+m\right)/2} \right)}{\sqrt{\pi } } \cdot \frac{K_{\left(1+{m/2} \right)} \left(\alpha \right)}{K_{1} \left(\alpha \right)} v_{0}^{m}.
\end{split}
\end{equation} 
The \textit{P} distribution can be easily obtained from Eqs. \eqref{ZEqnNum757414} and \eqref{ZEqnNum864132}, 
\begin{equation} 
\label{ZEqnNum229576} 
P\left(x\right)=\frac{e^{-\sqrt{\alpha ^{2} +{x/v_{0}^{2} } } } }{2\alpha v_{0} K_{1} \left(\alpha \right)\sqrt{x} } ,        \end{equation} 
where $P\left(x\right)$ approaches a Chi-square distribution of a single DoF (Degree of Freedom) when $\alpha \to \infty $. The \textit{m}th order moments and generalized kurtosis for \textit{P} distribution are,
\begin{equation} 
\label{eq:28} 
\begin{split}
M_{P} \left(m\right)&=\int _{0}^{\infty }P\left(v\right)v^{m}  dv\\
&=\frac{K_{1+m} \left(\alpha \right)}{K_{1} \left(\alpha \right)} \frac{\Gamma \left(m+{1/2} \right)}{\sqrt{\pi } } \left(2\alpha v_{0}^{2} \right)^{m},
\end{split}
\end{equation} 
\begin{equation} 
\label{eq:29} 
\begin{split}
K_{P} \left(m\right)&=\frac{M_{P} \left(m\right)}{\left(M_{P} \left(2\right)\right)^{{m/2} } } \\
&=\frac{K_{1+m} \left(\alpha \right)}{K_{1} \left(\alpha \right)} \left(\frac{K_{1} \left(\alpha \right)}{K_{3} \left(\alpha \right)} \right)^{{m/2} } \frac{\Gamma \left(m+{1/2} \right)}{\sqrt{\pi } } \left(\frac{2}{\sqrt{3} } \right)^{m} .   
\end{split}
\end{equation} 

With the help from Eqs. \eqref{ZEqnNum803629} and \eqref{ZEqnNum773894}, we can find the equation for \textit{H} distribution,
\begin{equation} 
\label{ZEqnNum273467} 
\int _{0}^{\infty }H\left(\sigma _{v}^{2} \right)e^{-\sigma ^{2} t} d\sigma _{v}^{2}  =\frac{K_{1} \left(\alpha \sqrt{1+2v_{0}^{2} t} \right)}{K_{1} \left(\alpha \right)\sqrt{1+2v_{0}^{2} t} } .       
\end{equation} 
The \textit{H} distribution is related to the dimensionless halo mass function $f\left(\nu \right)$, as we will show in Section \ref{sec:3}. Two limiting situations can be identified for \textit{H} distribution from Eq. \eqref{ZEqnNum803629},
\begin{equation}
H\left(x=\sigma _{v}^{2} \right)=\frac{e^{{-x/\sigma _{0}^{2} } } }{\sigma _{0}^{2} } \quad \textrm{for} \quad n=0 \quad \textrm{and} \quad \alpha \to 0
\label{ZEqnNum730411}
\end{equation}
\begin{equation}
H\left(x=\sigma _{v}^{2} \right)=\delta \left(x\right) \quad \textrm{for} \quad n=-2 \quad \textrm{and} \quad \alpha \to \infty.     
\label{eq:32}
\end{equation}

\noindent Obviously, $H(\sigma _{v}^{2})=\delta (\sigma _{v}^{2})$ means only one size of halo with the smallest virial dispersion $\sigma _{v}^{2} =0$ exists for short-range interaction system with $n=-2$. For system with $n=-1$, $H(\sigma _{v}^{2})$ is expected to be between two limiting situations.

This Section summarizes relevant limiting probability distributions and their relations. This will be used to provide insights into the halo mass function in next Section. Table \ref{tab:1} listed the relevant parameters and distributions for different potential exponent \textit{n}, along with the statistical properties of distributions in Table \ref{tab:A1}.

\begin{table*}
\caption{Parameters and distributions for some typical potential exponents \textit{n}}
\begin{tabular}{p{0.2in}p{0.2in}p{1in}p{0.6in}p{0.6in}p{0.6in}p{0.8in}p{0.7in}p{0.8in}} \hline 
$n$ & $\beta $ & $\alpha $ & $v_{0}^{2} $ & $\left\langle \sigma _{h}^{2} \right\rangle $ & $\left\langle \sigma _{v}^{2} \right\rangle $ & $X\left(v\right)$ & $H\left(x=\sigma _{v}^{2} \right)$ & $P\left(x=v^{2} \right)$ \\ \hline 
0 & 1 & 0 & $\frac{\sigma _{0}^{2} }{2} $ & 0 & $\sigma _{0}^{2} $ & $\frac{e^{{-\sqrt{2} v/\sigma _{0}^{} } } }{\sqrt{2} \sigma _{0}^{} } $ & $\frac{e^{{-x/\sigma _{0}^{2} } } }{\sigma _{0}^{2} } $ & $\frac{e^{{-\sqrt{2x} /\sigma _{0}^{} } } }{\sigma _{0}^{} \sqrt{2x} } $ \\ \hline 
-1 & $\frac{3}{2} $ & $\frac{K_{1} \left(\alpha \right)}{K_{2} \left(\alpha \right)} =\frac{\left\langle \sigma _{h}^{2} \right\rangle }{\sigma _{0}^{2} } $ & $\frac{\sigma _{0}^{2} K_{1} \left(\alpha \right)}{\alpha K_{2} \left(\alpha \right)} $ & $\sim \frac{\sigma _{0}^{2} }{2} $ & $\sim \frac{\sigma _{0}^{2} }{2} $ & $\frac{e^{-\sqrt{\alpha ^{2} +\left({v/v_{0} } \right)^{2} } } }{2\alpha v_{0} K_{1} \left(\alpha \right)} $ &  & $\frac{e^{-\sqrt{\alpha ^{2} +{x/v_{0}^{2} } } } }{2\alpha v_{0}^{} K_{1} \left(\alpha \right)\sqrt{x} } $ \\ \hline 
-2 & 3 & $\infty $ & 0 & $\sigma _{0}^{2} $ & 0 & $\frac{e^{{-v^{2} /2\sigma _{0}^{2} } } }{\sqrt{2\pi } \sigma _{0}^{} } $ & $\delta \left(x\right)$ & $\frac{e^{{-x/2\sigma _{0}^{2} } } }{\sigma _{0}^{} \sqrt{2\pi x} } $ \\ \hline 
\end{tabular}
\label{tab:2}
\end{table*}

\subsection{\texorpdfstring{$H_{\infty }$}{} and \texorpdfstring{$J_{\infty}$}{} distributions for \texorpdfstring{$\sigma _{h}^{2}=0$}{} and \texorpdfstring{$\sigma ^{2} =\sigma _{v}^{2}$}{} (large halos)}
\label{sec:2.3}
We first consider an extreme case with $\sigma _{h}^{2} =0$ and $\sigma ^{2} =\sigma _{v}^{2} $ that is relevant to large halos. The virial dispersion of large halos is dominant over halo velocity dispersion with $\sigma _{v}^{2} \gg \sigma _{h}^{2} $ and $\sigma ^{2} \approx \sigma _{v}^{2} $. For this limiting case, the analytical solution of \textit{H} distribution can be found from Eq. \eqref{ZEqnNum273467}, 
\begin{equation} 
\label{ZEqnNum813447} 
H_{\infty } \left(\sigma _{v}^{2} \right)=\frac{1}{2\alpha v_{0}^{2} K_{1} \left(\alpha \right)} \cdot \exp \left[-\frac{\alpha }{2} \left(\frac{\sigma _{v}^{2} }{\alpha v_{0}^{2} } +\frac{\alpha v_{0}^{2} }{\sigma _{v}^{2} } \right)\right].      
\end{equation} 
The statistical properties of the $H_{\infty } $ distribution can be easily computed and listed here. The moment-generating function for $H_{\infty } $ distribution is:
\begin{equation} 
\label{eq:34} 
MGF_{H_{\infty } } \left(t\right)=\frac{K_{1} \left(\alpha \sqrt{1-2v_{0}^{2} t} \right)}{K_{1} \left(\alpha \right)\sqrt{1-2v_{0}^{2} t} } .        
\end{equation} 
The \textit{m}th moment of $H_{\infty } $ distribution and generalized kurtosis are
\begin{equation} 
\label{eq:35} 
M_{H_{\infty } } \left(m\right)=\frac{K_{1+m} \left(\alpha \right)}{K_{1} \left(\alpha \right)} \left(\alpha v_{0}^{2} \right)^{m} ,          
\end{equation} 
and
\begin{equation} 
\label{eq:36} 
K_{H_{\infty } } \left(m\right)=\frac{K_{1+m} \left(\alpha \right)}{K_{1} \left(\alpha \right)} \left(\frac{K_{1} \left(\alpha \right)}{K_{3} \left(\alpha \right)} \right)^{{m/2} } .        
\end{equation} 
The Shannon entropy of the $H_{\infty } $ distribution is
\begin{equation} 
\label{eq:37} 
\begin{split}
S_{H_{\infty } }&=-\int _{-\infty }^{\infty }H\left(\sigma _{v}^{2} \right)\ln \left(H\left(\sigma _{v}^{2} \right)\right) d\sigma _{v}^{2}\\
&=\alpha \frac{K_{2} \left(\alpha \right)}{K_{1} \left(\alpha \right)} -1+\ln \left(2\alpha v_{0}^{2} K_{1} \left(\alpha \right)\right). 
\end{split}
\end{equation} 
Table \ref{tab:A1} lists statistical properties of $H_{\infty}$ and \textit{P} distributions. 

Now, let's find the $J_{\infty } $ distribution for the number of halos corresponding to $H_{\infty } $ distribution. We first assume a power-law for the mass-dispersion relation from the virial theorem, 
\begin{equation}
\label{ZEqnNum929629} 
n_{p} \left(\sigma _{v}^{2} \right)=\mu \bar{N}\left({\sigma _{v}^{2} /v_{0}^{2} } \right)^{\beta } ,         
\end{equation} 
where $\mu $ is a normalization constant and exponent $\beta ={3/\left(3+n\right)} $ (Eq. \eqref{ZEqnNum386172}). The $J_{\infty}$ distribution can be found using Eq. \eqref{ZEqnNum639447},
\begin{equation} 
\label{eq:39} 
J_{\infty } \left(\sigma _{v}^{2} \right)=H_{\infty } \left(\sigma _{v}^{2} \right)\frac{\bar{N}}{n_{h} \left(\sigma _{v}^{2} \right)} =\frac{H_{\infty } \left(\sigma _{v}^{2} \right)}{\mu \left({\sigma _{v}^{2} /v_{0}^{2} } \right)^{\beta } } .       
\end{equation} 
The normalization condition $\int _{0}^{\infty }J_{\infty } \left(\sigma _{v}^{2} \right) d\sigma _{v}^{2} =1$ requires, 
\begin{equation} 
\label{eq:40} 
\mu =\alpha ^{-\beta } \frac{K_{\beta -1} \left(\alpha \right)}{K_{1} \left(\alpha \right)} .           
\end{equation} 
The final expression for $J_{\infty } $ distribution is,
\begin{equation} 
\label{eq:41} 
J_{\infty } \left(\sigma _{v}^{2} \right)=\frac{1}{2\alpha v_{0}^{2} K_{\beta -1} \left(\alpha \right)} \left(\frac{\alpha v_{0}^{2} }{\sigma _{v}^{2} } \right)^{\beta } \exp \left[-\frac{\alpha }{2} \left(\frac{\sigma _{v}^{2} }{\alpha v_{0}^{2} } +\frac{\alpha v_{0}^{2} }{\sigma _{v}^{2} } \right)\right] 
\end{equation} 
and the mass-dispersion relation is (from Eq. \eqref{ZEqnNum929629})
\begin{equation} 
\label{eq:42} 
n_{p} \left(\sigma _{v}^{2} \right)=\bar{N}\frac{K_{\beta -1} \left(\alpha \right)}{K_{1} \left(\alpha \right)} \left(\frac{\sigma _{v}^{2} }{\alpha v_{0}^{2} } \right)^{\beta } .        
\end{equation} 
Similarly, the moment-generating function for $J_{\infty } $ distribution is:
\begin{equation} 
\label{eq:43} 
MGF_{J_{\infty } } \left(t\right)=\frac{K_{\beta -1} \left(\alpha \sqrt{1-2v_{0}^{2} t} \right)}{K_{\beta -1} \left(\alpha \right)\left(\sqrt{1-2v_{0}^{2} t} \right)^{1-\beta } } .        
\end{equation} 
The \textit{m}th moment of $J_{\infty}$ distribution and generalized kurtosis are
\begin{equation} 
\label{eq:44} 
M_{J_{\infty } } \left(m\right)=\frac{K_{m+1-\beta } \left(\alpha \right)}{K_{1-\beta } \left(\alpha \right)} \left(\alpha v_{0}^{2} \right)^{m} ,        
\end{equation} 
\begin{equation} 
\label{eq:45} 
K_{J_{\infty } } \left(m\right)=\frac{K_{1+m-\beta } \left(\alpha \right)}{K_{1-\beta } \left(\alpha \right)} \left(\frac{K_{1-\beta } \left(\alpha \right)}{K_{3-\beta } \left(\alpha \right)} \right)^{{m/2} } .        
\end{equation} 

\subsection{\texorpdfstring{$H_{\infty }$}{} distribution from maximum entropy principle }
\label{sec:2.4}
Interestingly, $H_{\infty } $ distribution (Eq. \eqref{ZEqnNum813447}) can be obtained directly using the maximum entropy principle without resorting to \textit{X} distribution. We can show that $H_{\infty } $ distribution is actually a maximum entropy distribution satisfying three constraints,
\begin{equation} 
\label{ZEqnNum371879} 
\int _{0}^{\infty }H_{\infty } \left(\sigma _{v}^{2} \right)d\sigma _{v}^{2}  =1,          
\end{equation} 
\begin{equation} 
\label{eq:47} 
\int _{0}^{\infty }H_{\infty } \left(\sigma _{v}^{2} \right)\sigma _{v}^{2} d\sigma _{v}^{2}  =\left\langle \sigma _{v}^{2} \right\rangle ,         
\end{equation} 
\begin{equation} 
\label{ZEqnNum490747} 
\int _{0}^{\infty }J_{\infty } \left(\sigma _{v}^{2} \right)d\sigma _{v}^{2}  =\int _{0}^{\infty }\frac{H_{\infty } \left(\sigma _{v}^{2} \right)}{\mu \left({\sigma _{v}^{2} /v_{0}^{2} } \right)^{\beta } } d\sigma _{v}^{2}  =1.        
\end{equation} 
The third constraint is from the normalization condition of $J_{\infty}$ distribution that must be satisfied by $H_{\infty}$ distribution. The entropy functional of $H_{\infty}$ distribution can be constructed as:
\begin{equation} 
\label{ZEqnNum758060} 
\begin{split}
S\left[H_{\infty } \left(\sigma _{v}^{2} \right)\right]=&-\int _{0}^{\infty }H_{\infty } \left(\sigma _{v}^{2} \right)\ln H_{\infty } \left(\sigma _{v}^{2} \right)d\sigma _{v}^{2} \\
& +\lambda _{1} \left(\int _{0}^{\infty }H_{\infty } \left(\sigma _{v}^{2} \right)d\sigma _{v}^{2} - 1\right) \\
&+\lambda _{2} \left(\int _{0}^{\infty }H_{\infty } \left(\sigma _{v}^{2} \right)\sigma _{v}^{2} d\sigma _{v}^{2} - \left\langle \sigma _{v}^{2} \right\rangle \right)\\
&+\lambda _{3} \left(\int _{0}^{\infty }\frac{H_{\infty } \left(\sigma _{v}^{2} \right)}{\mu \left({\sigma _{v}^{2} /v_{0}^{2} } \right)^{\beta } } d\sigma _{v}^{2} - 1\right).
\end{split}
\end{equation} 
The variation of entropy functional (Eq. \eqref{ZEqnNum758060}) with respect to the $H_{\infty } $ distribution should vanish, which leads to a maximum entropy distribution
\begin{equation} 
\label{ZEqnNum234675} 
H_{\infty } \left(\sigma _{v}^{2} \right)=e^{\lambda _{1} -1} \exp \left(\lambda _{2} \sigma _{v}^{2} +\frac{\lambda _{3} }{\mu } \left(\frac{v_{0}^{2} }{\sigma _{v}^{2} } \right)^{\beta } \right).   \end{equation} 
Obviously, with three Lagrange multipliers
\begin{equation}
\lambda _{1} =1-\ln \left(2\alpha v_{0}^{2} K_{1} \left(\alpha \right)\right), \quad \lambda _{2} =-{1}/{\left(2v_{0}^{2} \right)}, \quad \lambda _{3} =-{\mu \alpha ^{2}}/{2},    
\label{eq:51}
\end{equation}
\noindent and $\beta =1$ or potential exponent $n=0$, Eq. \eqref{ZEqnNum234675} reduces to the $H_{\infty } $ distribution in Eq. \eqref{ZEqnNum813447}. The $H_{\infty } $ distribution is essentially a maximum entropy distribution satisfying three constraints (Eqs. \eqref{ZEqnNum371879}-\eqref{ZEqnNum490747}) with $\beta =1$ or equivalently $n=0$. With $n=0$ and $\alpha =0$, the $H_{\infty } $ distribution in Eq. \eqref{ZEqnNum813447} also reduces to the \textit{H} distribution in Eq. \eqref{ZEqnNum730411}. i.e. the \textit{H} distribution for $n=0$ is also a maximum entropy distribution. For large halos with $\sigma _{v}^{2} \to \infty $, $H_{\infty } $ distribution in Eq. \eqref{ZEqnNum813447} reduces to 
\begin{equation} 
\label{eq:52} 
H_{\infty } \left(\sigma _{v}^{2} \to \infty \right)=\frac{1}{2\alpha v_{0}^{2} K_{1} \left(\alpha \right)} \cdot \exp \left(-\frac{\sigma _{v}^{2} }{2v_{0}^{2} } \right),       
\end{equation} 
which is an exponential distribution for large halos with mean $2v_{0}^{2}$.

\section{Halo mass function from maximum entropy distributions}
\label{sec:3}
\subsection{The simulation data}
\label{sec:3.1}
The numerical data used in this paper are public available and generated from \textit{N}-body simulations carried out by the Virgo consortium. A comprehensive description of the data can be found in \citep{Frenk:2000-Public-Release-of-N-body-simul,Jenkins:1998-Evolution-of-structure-in-cold}. The current study was carried out using the simulation runs with Ω = 1 and the standard CDM power spectrum (SCDM) to focus on the matter-dominant gravitational collapse of collisionless particles. The same set of data has been widely used in a number of studies from clustering statistics \citep{Jenkins:1998-Evolution-of-structure-in-cold} to formation of cluster halos in large scale environment \citep{Colberg:1999-Linking-cluster-formation-to-l}, and test of models for halo abundances and mass functions \citep{Sheth:2001-Ellipsoidal-collapse-and-an-im}. The friends-of-friends algorithm (FOF) was used to identify halos that depends on just one parameter \textit{b}, which defines the linking length $b\left({N_{p} /V} \right)^{{-1/3} } $. Halos are identified with a linking length $b=0.2$ in this work. Some key parameters of N-body simulations are listed in Table \ref{tab:1}.

Two relevant datasets from this N-boby simulation, i.e. halo-based and correlation-based statistics of dark matter flow, can be found at Zenodo.org  \citep{Xu:2022-Dark_matter-flow-dataset-part1, Xu:2022-Dark_matter-flow-dataset-part2}, along with the accompanying presentation slides, "A comparative study of dark matter flow \& hydrodynamic turbulence and its applications" \citep{Xu:2022-Dark_matter-flow-and-hydrodynamic-turbulence-presentation}. All data files are also available on GitHub \citep{Xu:Dark_matter_flow_dataset_2022_all_files}.

\begin{table}
\caption{Numerical parameters of N-body simulation}
\begin{tabular}{p{0.25in}p{0.05in}p{0.05in}p{0.05in}p{0.05in}p{0.05in}p{0.4in}p{0.1in}p{0.4in}p{0.4in}} 
\hline 
Run & $\Omega_{0}$ & $\Lambda$ & $h$ & $\Gamma$ & $\sigma _{8}$ & \makecell{L\\(Mpc/h)} & $N$ & \makecell{$m_{p}$\\$M_{\odot}/h$} & \makecell{$l_{soft}$\\(Kpc/h)} \\ 
\hline 
SCDM1 & 1.0 & 0.0 & 0.5 & 0.5 & 0.51 & \centering 239.5 & $256^{3}$ & 2.27$\times 10^{11}$ & \makecell{\centering 36} \\ 
\hline 
\end{tabular}
\label{tab:1}
\end{table}

\subsection{Halo virial dispersion and halo velocity dispersion}
\label{sec:3.2}
\begin{figure}
\includegraphics*[width=\columnwidth]{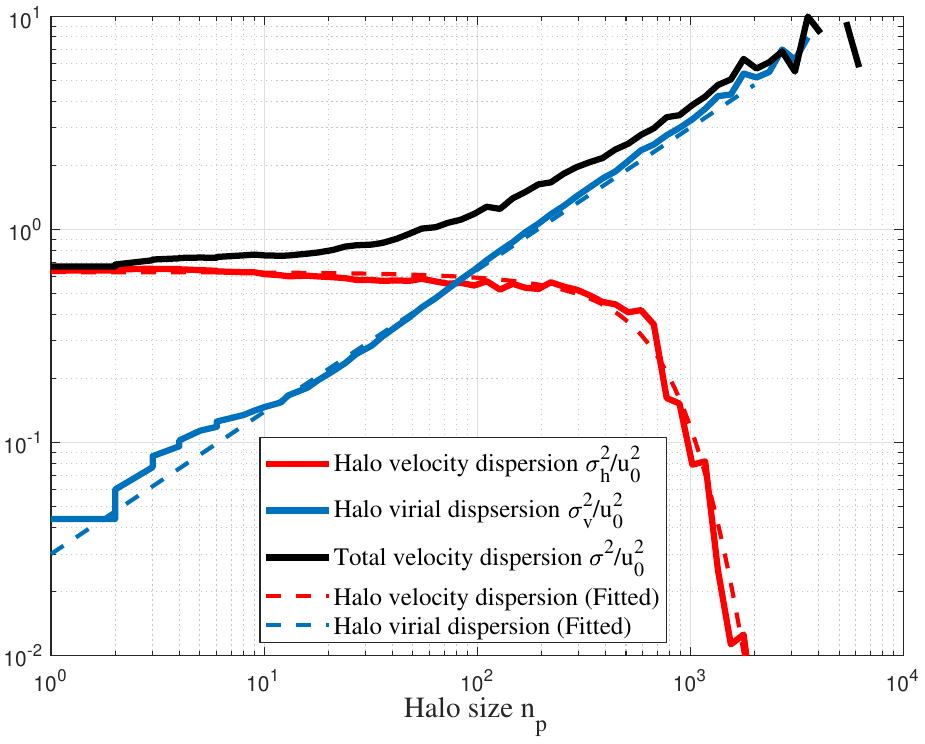}
\caption{The variation of halo virial dispersion and velocity dispersion (halo and group temperature) with halo size from a \textit{N}-body simulation. Obviously, $\sigma _{v}^{2} \gg \sigma _{h}^{2} $ for massive halos and $\sigma _{v}^{2} \ll \sigma _{h}^{2} $ for small halos. The halo velocity dispersion increases with halo size as $\sigma _{v}^{2} \propto m_{h}^{{2/3} } $. The halo velocity dispersion for halo groups slowly decreases with the halo size up to $m_{h} \approx 600n_{p} $, followed by a sharp drop to zero for large halos. To a first order approximation, one may assume that $\sigma _{h}^{2} $ is independent of the halo size.}
\label{fig:2}
\end{figure}

Equation \eqref{ZEqnNum805703} defines two temperatures, namely halo temperature from virial dispersion $\sigma _{v}^{2} $ (average temperature of all halos in the same group) and a halo group temperature from the motion of halos (velocity dispersion $\sigma _{h}^{2} $).  Figure \ref{fig:2} plots the variation of two velocity dispersions and the total dispersion $\sigma _{}^{2} $ with the halo size $n_{p} $ using simulation data at redshift \textit{z}=0. In Fig. \ref{fig:2}, two dispersions have very different dependence with the halo size. Obviously, $\sigma _{v}^{2} \gg \sigma _{h}^{2}$ for massive and hot halos and $\sigma _{v}^{2} \ll \sigma _{h}^{2} $ for small and cold halos. 

The (large) halo virial dispersion $\sigma _{v}^{2} \propto a^{-1} \left(m_{h} \right)^{{2/3} } $ and a convenient fitting formula is provided in \citep{Bryan:1998-Statistical-properties-of-X-ra}. The halo velocity dispersion $\sigma _{h}^{2} $ for halo groups slowly decreases with the halo size up to $m_{h} \approx 500m_{p}$, followed by a sharp decrease to zero for large halos. For a first order approximation, one may assume that $\sigma _{h}^{2} $ is independent of the halo size. Hence, $\sigma _{h}^{2} \left(m_{h} \right)=\sigma _{hc}^{2} $ is a constant for all halo groups of different sizes, where $\sigma _{hc}^{2} $ is the background temperature of entire system. The mean halo velocity dispersion $\bar{\sigma }_{h}^{2} $ is defined in Eq. \eqref{ZEqnNum435332} and $\sigma _{h0}^{2} =\sigma _{h}^{2} \left(m_{h} =0\right)$ is the largest halo velocity dispersion for smallest halo with $m_{h} =0$. From simulation data at \textit{z}=0, the average and largest halo velocity dispersion are found as \citep{Xu:2021-Inverse-and-direct-cascade-of-},
\begin{equation}
\bar{\sigma }_{h}^{2} =0.57u_{0}^{2} \quad \textrm{and} \quad \sigma _{h0}^{2} =0.65u_{0}^{2},       
\label{eq:53}
\end{equation}
\noindent where $u_{0}^{2} =354.6{km/s} $ is the one-dimensional velocity dispersion of the entire system at \textit{z}=0. A better fit to the simulation data can be obtained for both velocity dispersions at $z=0$ is,
\begin{equation} 
\label{eq:54} 
\sigma _{v}^{2} \left(m_{h} \right)=0.03n_{p}^{{2/3} } u_{0}^{2} =0.03\left({m_{h} /m_{p} } \right)^{{2/3} } u_{0}^{2} \end{equation} 
and
\begin{equation} 
\label{ZEqnNum933840} 
\sigma _{h}^{2} \left(m_{h} \right)=0.375\left[1-\tanh \left(\frac{{m_{h} /m_{p} } -500}{600} \right)\right]u_{0}^{2}. \end{equation}

A characteristic mass scale $m_{h}^{*} $ can be defined at which the halo virial dispersion equals the mean halo velocity dispersion,
\begin{equation} 
\label{ZEqnNum523382} 
\bar{\sigma }_{h}^{2} \left(a\right)=\left\langle \sigma _{h}^{2} \left(m_{h} ,a\right)\right\rangle =\sigma _{v}^{2} \left(m_{h}^{*} ,a\right),        
\end{equation} 
where \textit{a} is the scale factor and $m_{h}^{*} \left(a\right)$ is a mass scale that is about $2\times 10^{13} {M_{\odot } /h} $ at \textit{z}=0, or an equivalent halo size of $n_{p}^{*} \approx 80$. Halos with mass greater than the critical mass ($m_{h}^{} >m_{h}^{*} $) are hotter than the halo group ($\sigma _{v}^{2} >\sigma _{h}^{2} $). Large halos are much rarer compared to halos with mass smaller than the critical mass $m_{h}^{*} $.

\subsection{Halo mass functions from simulations and existing models}
\label{sec:3.3}
In this section, connections between \textit{H} distribution and halo mass function will be presented. With mean velocity dispersion $\bar{\sigma }_{h}^{2} $ in Eq. \eqref{ZEqnNum523382}, a dimensionless variable $\nu $ can be defined as,
\begin{equation} 
\label{ZEqnNum423968} 
v=\left(\frac{m_{h} }{m_{h}^{*} } \right)^{{2/3} } =\frac{\sigma _{v}^{2} \left(m_{h} \right)}{\sigma _{v}^{2} \left(m_{h}^{*} \right)} =\frac{\sigma _{v}^{2} \left(m_{h} \right)}{\bar{\sigma }_{h}^{2} } .        
\end{equation} 
From linear theory, the variable $\nu $ can be related to the variance $\sigma _{\delta }^{2} \left(m_{h} ,z\right)$ of the density fluctuation when smoothed by a tophat filter with a size of halo of mass $m_{h} $,
\begin{equation} 
\label{eq:58} 
\nu \propto v_{p}^{2} =\left[\frac{\delta _{c}^{} }{\sigma _{\delta }^{} \left(m_{h} ,z\right)} \right]^{2} ,          
\end{equation} 
where $\delta _{c}^{} =1.686$ is the critical density from spherical collapse model or two-body collapse model \citep{Xu:2021-A-non-radial-two-body-collapse} and $v_{p} $ is the peak height of halos. With Eq. \eqref{ZEqnNum423968}, the \textit{H} distribution can be equivalently transformed to a new distribution $f\left(v\right)$ 
\begin{equation} 
\label{ZEqnNum330225} 
f\left(v\right)=H\left(v\bar{\sigma }_{h}^{2} \right)\bar{\sigma }_{h}^{2} ,          
\end{equation} 
which is exactly the dimensionless halo mass function. In simulation, we first compute the \textit{H} distribution. All particles in the same halo group are given a virial dispersion $\sigma _{v}^{2} $ of that group and this operation is performed over all particles from all halos identified in the simulation. The \textit{H} distribution is just the fraction of particles with a given virial dispersion between $\left[\sigma _{v}^{2} ,\sigma _{v}^{2} +d\sigma _{v}^{2} \right]$. Halo mass function $f\left(v\right)$ can be obtained from \textit{H} distribution using Eq. \eqref{ZEqnNum330225}. 

\begin{figure}
\includegraphics*[width=\columnwidth]{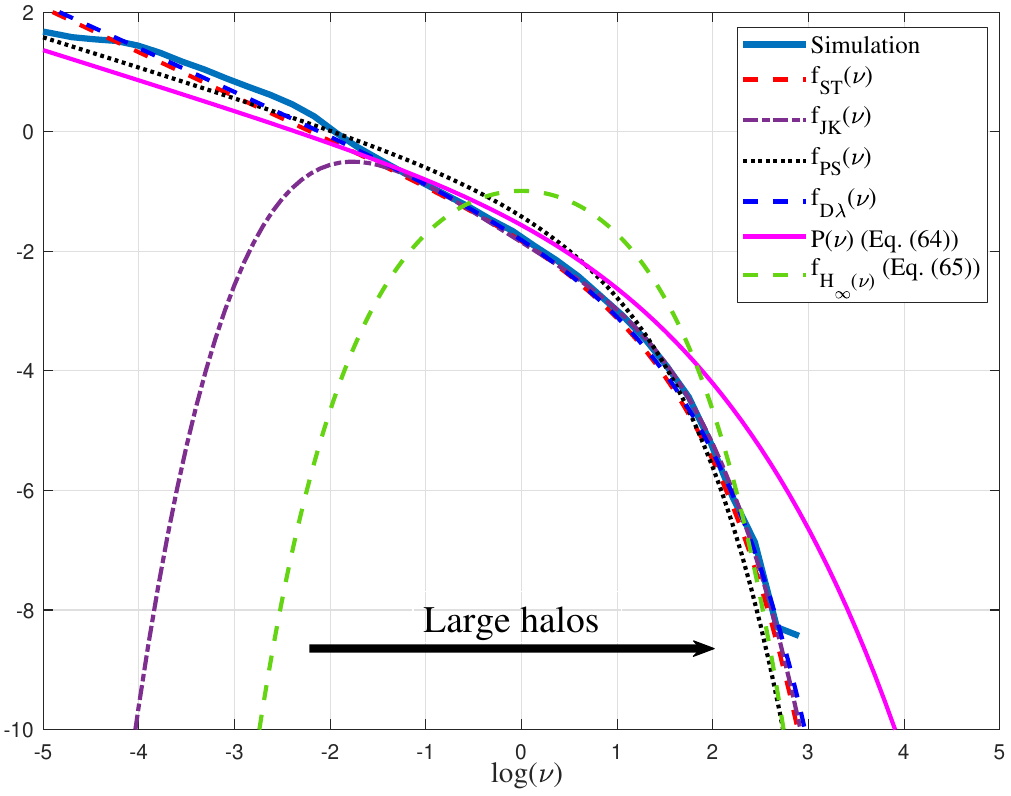}
\caption{The mass function obtained from a \textit{N}-body simulation from \textit{H} distribution (the solid blue line). The other four mass functions, i.e. PS, ST, JK and Double-$\lambdaup$ models, are also presented in the same figure. Both Double-$\lambdaup$ and ST models reasonably match the simulation results. The PS mass function underestimates the mass in large halos and overestimate the mass in small halos. The fitted JK mass function only matches the simulation for a certain range of halo mass. The \textit{P} and $H_{\infty } $ distributions are also plotted that approximate the PS mass function for small and large halos.}
\label{fig:3}
\end{figure}

The mass function obtained this way with $\bar{\sigma }_{h}^{2} =0.57u_{0}^{2} $ is presented in Fig. \ref{fig:3} as the solid line along with the PS, ST, JK and Double-$\lambdaup$ mass functions from Eqs. \eqref{ZEqnNum333244}-\eqref{ZEqnNum982412}. Both Double-$\lambdaup$ and ST models reasonably match the simulation results. The PS mass function underestimates the mass in large halos and overestimate the mass in small halos. The fitted JK mass function matches the simulation only for a given range of halo mass that is used for fitting. The \textit{P} and $H_{\infty } $ distributions are also plotted in the same figure that approximate the PS mass function for small and large halos. This will be discussed in next section. 

\subsection{Halo mass function for small halos with \texorpdfstring{$\sigma _{v}^{2} =0$ and $\sigma ^{2} =\sigma _{h}^{2} $}{}}
\label{sec:3.4}
The limiting case for large halos was discussed in Section \ref{sec:2.3}. For the other limiting case, i.e. small halos with virial dispersion $\sigma _{v}^{2} \to 0$, the \textit{H} distribution can be approximated by a \textit{P(x)} distribution with $x=\sigma _{v}^{2} $, i.e. $H(x)\approx P(x)$. Let's revisit the relation between \textit{P} and \textit{H} distributions (Eq. \eqref{ZEqnNum757414}). If term 1 in Eq. \eqref{ZEqnNum757414} for \textit{P} distribution, i.e. a one DoF Chi-square distribution, can be approximated by a direct delta function for small $\sigma ^{2} $, $\sigma _{v}^{2} $ and $x$, 
\begin{equation} 
\label{eq:60} 
\frac{1}{\sqrt{2\pi x} \sigma } e^{-{x/2\sigma ^{2} } } \approx \delta \left(x-\sigma _{v}^{2} \right),          
\end{equation} 
such that \textit{H} can be approximated by \textit{P} as (using Eqs. \eqref{ZEqnNum757414} and \eqref{ZEqnNum229576})
\begin{equation}
H\left(x=\sigma _{v}^{2} \right)\approx P\left(x=v^{2} \right)=\frac{e^{-\alpha } }{2\alpha v_{0} K_{1} \left(\alpha \right)\sqrt{x} } \exp \left(-\frac{x}{2\alpha v_{0}^{2} } \right) 
\label{ZEqnNum789799}
\end{equation}
\noindent for  $\sigma _{v}^{2} \to 0$. This can be interpreted as: most particles with a small $v^{2} $ reside in small halos with a small virial dispersion $\sigma _{v}^{2} $. In the limiting situation, all particles in the smallest halos with $\sigma _{v}^{2} \to 0$ will also have $v^{2} =0$ if the halo velocity dispersion $\sigma _{h}^{2} $ also approaches zero. Therefore, the fraction of particles with a square speed $v^{2} \to 0$ is the same as the fraction of particles with $\sigma _{v}^{2} \to 0$ if $\sigma _{h}^{2} \to 0$ for small halos. This is of course a quite crude approximation, as we know that $\sigma _{h}^{2} \left(m_{h} \right)\to \sigma _{h0}^{2} \ne 0$ for small halos with $m_{h} \to 0$ (Fig. \ref{fig:2}). 

Nonetheless, the \textit{H} distribution for small halos can be reasonably approximated by the \textit{P} distribution for small halos, where particle velocity follows the Gaussian core of the \textit{X} distribution (Eq. \eqref{ZEqnNum174426}) with a variance of $\alpha v_{0}^{2} $. The final distribution based on Eq. \eqref{ZEqnNum789799} is
\begin{equation} 
\label{ZEqnNum988300} 
H_{s} \left(\sigma _{v}^{2} \right)=\frac{1}{\sqrt{2\pi \alpha v_{0}^{2} \sigma _{v}^{2} } } \exp \left(-\frac{\sigma _{v}^{2} }{2\alpha v_{0}^{2} } \right),       
\end{equation} 
a (one degree of freedom) Chi-square distribution with a mean of $\alpha v_{0}^{2} $ that is consistent with PS mass function in Eq. \eqref{ZEqnNum333244}. The dimensionless mass function can be found using Eqs. \eqref{ZEqnNum988300} and \eqref{ZEqnNum330225},
\begin{equation} 
\label{ZEqnNum948455} 
f_{H_{s} } \left(\nu \right)=\frac{1}{\sqrt{2\pi \gamma \nu } } \exp \left(-\frac{\nu }{2\gamma } \right),         
\end{equation} 
where $\gamma ={\alpha v_{0}^{2} /\bar{\sigma }_{h}^{2} }$ is a dimensionless parameter of order unity. Clearly, the mass function in Eq. \eqref{ZEqnNum948455} reduces to PS mass function with $\gamma \approx {\left\langle \sigma _{v}^{2} \right\rangle /\bar{\sigma }_{h}^{2} } \approx 1$. 

For comparison, \textit{P} distribution can be normalized as (Eqs. \eqref{ZEqnNum229576})
\begin{equation} 
\label{eq:64} 
P\left(\nu =\frac{v^{2} }{\bar{\sigma }_{h}^{2} } \right)=\frac{e^{-\sqrt{\alpha ^{2} +{\alpha \nu /\gamma } } } }{2K_{1} \left(\alpha \right)\sqrt{\alpha \gamma \nu } } .        
\end{equation} 
Similarly, the $H_{\infty } $ solution we obtained in Section \ref{sec:2.2} (Eq. \eqref{ZEqnNum813447}) for $\sigma _{h}^{2} =0$ can be rewritten as
\begin{equation} 
\label{ZEqnNum229053} 
f_{H_{\infty } } \left(\nu \right)=\frac{1}{2\gamma K_{1} \left(\alpha \right)} \cdot \exp \left[-\frac{\alpha }{2} \left(\frac{\nu }{\gamma } +\frac{\gamma }{\nu } \right)\right].        
\end{equation} 
Figure \ref{fig:3} presents the $P\left(\nu \right)$ and $f_{H_{\infty } } \left(\nu \right)$ in the same plot with $\alpha =1.33$ and $\gamma =1$ from simulation. As expected, the distribution $P\left(\nu \right)$ approximates the PS mass function for small halos and $f_{H_{\infty } } \left(\nu \right)$ approximates the mass function for large halos. A better fitting may be obtained by adjusting the values of parameters $\alpha $ and $\gamma $.

\subsection{Halo mass function from maximum entropy distributions}
\label{sec:3.5}
With two limiting situations discussed in Section \ref{sec:2.3} for large halos and Section \ref{sec:3.4} for small halos, now let's revisit the equation for \textit{H} distribution to have more insights. The dimensionless mass function $f\left(\nu \right)$ should satisfy (from Eqs. \eqref{ZEqnNum273467} and  \eqref{ZEqnNum862024})  
\begin{equation} 
\label{ZEqnNum883658} 
\int _{0}^{\infty }f\left(\nu \right) e^{-\left(\nu +v_{h} \right)t} d\nu =\frac{K_{1} \left(\alpha \sqrt{1+2{\gamma t/\alpha } } \right)}{K_{1} \left(\alpha \right)\sqrt{1+2{\gamma t/\alpha } } } ,       
\end{equation} 
and
\begin{equation} 
\label{eq:67} 
\int _{0}^{\infty }f\left(\nu \right) \left(\nu +\nu _{h} \right)^{{m/2} } d\nu =\gamma ^{{m/2} } \frac{K_{\left(1+{m/2} \right)} \left(\alpha \right)}{K_{1} \left(\alpha \right)} ,       
\end{equation} 
where $\nu _{h} ={\sigma _{h}^{2} /\bar{\sigma }_{h}^{2} } $ is a normalized halo velocity dispersion. 

In principle, the dimensionless mass function $f\left(\nu \right)$ can be obtained by solving Eq. \eqref{ZEqnNum883658} for a given model of $\nu _{h}$, parameters $\alpha$, and $\gamma$. A special case $\nu _{h} =0$ leads to the solution of mass function $f\left(\nu \right)$ in Eq. \eqref{ZEqnNum229053}, i.e. the limiting case for large halos. However, it is challenging to solve Eq. \eqref{ZEqnNum883658} for function $\nu _{h} \ne 0$, where no closed-form solution can be available. A constant halo velocity dispersion can be a good approximation such that $\sigma _{h}^{2} \left(m_{h} \right)\approx \sigma _{hc}^{2} $. Here $\sigma _{hc}^{2} $ is an effective halo velocity dispersion,
\begin{equation} 
\label{ZEqnNum385449} 
\nu _{h} \left(m_{h} \right)=\frac{\sigma _{h}^{2} \left(m_{h} \right)}{\bar{\sigma }_{h}^{2} } \approx \lambda =\frac{\sigma _{hc}^{2} }{\bar{\sigma }_{h}^{2} } ,         
\end{equation} 
where $\lambda \approx 1$ is expected. For a constant normalized dispersion $\nu _{h} $, an approximation for Laplace transform of $f\left(\nu \right)$ from Eqs. \eqref{ZEqnNum883658},
\begin{equation} 
\label{eq:69} 
\int _{0}^{\infty }f\left(\nu \right) e^{-\nu t} d\nu \approx \frac{K_{1} \left(\alpha \sqrt{1+2{\gamma t/\alpha } } \right)}{K_{1} \left(\alpha \right)\sqrt{1+2{\gamma t/\alpha } } } e^{\lambda t} ,       
\end{equation} 
where the mass function $f\left(\nu \right)$ is fully determined by three dimensionless parameters ($\alpha $, $\gamma $ and $\lambda $). 

Among three parameters, $\alpha $ is a shape parameter that is only dependent on the potential exponent \textit{n.} Parameters $\gamma $ and $\lambda $ are dimensionless constants that should be independent of the redshift \textit{z}. Therefore, the dimensionless halo mass function $f\left(\nu \right)$ should be independent of redshift once the statistically steady state is reached. The mass function $f\left(\nu \right)$ should also maximize the system entropy because it can be directly related to \textit{X} distribution that maximizes the system entropy (Eq. \eqref{ZEqnNum803629}). 

For the purpose of comparison, we also present the moment functions for dimensionless PS (Eq. \eqref{ZEqnNum333244}), ST (Eq. \eqref{ZEqnNum971057}) and Double-$\lambdaup$ (\eqref{ZEqnNum982412}) mass functions, 
\begin{equation} 
\label{ZEqnNum181175} 
\int _{0}^{\infty }f_{PS} \left(\nu \right) \nu ^{n} d\nu =2^{n} \frac{\Gamma \left({1/2} +n\right)}{\sqrt{\pi } } ,        
\end{equation} 
\vspace*{-15pt}
\begin{equation} 
\label{ZEqnNum244876} 
\int _{0}^{\infty }f_{ST} \left(\nu \right) \nu ^{n} d\nu =\left(\frac{2}{q} \right)^{2} \frac{\Gamma \left({1/2} +n\right)+2^{-p} \Gamma \left({1/2} +n-p\right)}{\Gamma \left({1/2} \right)+2^{-p} \Gamma \left({1/2} -p\right)} , 
\end{equation} 
\vspace*{-15pt}
\begin{equation} 
\label{eq:72} 
\int _{0}^{\infty }f_{D\lambda } \left(\nu \right) \nu ^{n} d\nu =\frac{\left(4\eta _{0} \right)^{n} \Gamma \left({q/2} +n\right)}{\Gamma \left({q/2} \right)} .        
\end{equation} 
The Laplace transform of three mass functions are
\begin{equation} 
\label{eq:73} 
\int _{0}^{\infty }f_{PS} \left(\nu \right) e^{-\nu t} d\nu =\frac{1}{\sqrt{1+2t} } ,         
\end{equation} 
\vspace*{-15pt}
\begin{equation} 
\label{eq:74} 
\int _{0}^{\infty }f_{ST} \left(\nu \right) e^{-\nu t} d\nu =\frac{\sqrt{q} }{\sqrt{q+2t} } \frac{\sqrt{\pi } +\Gamma \left({1/2} -p\right)\left({1/2} +{t/q} \right)^{p} }{\sqrt{\pi } +2^{-p} \Gamma \left({1/2} -p\right)} ,     
\end{equation} 
\vspace*{-15pt}
\begin{equation} 
\label{eq:75} 
\int _{0}^{\infty }f_{D\lambda } \left(\nu \right) e^{-\nu t} d\nu =\frac{1}{\left(1+4\eta _{0} t\right)^{{q/2} } } .        
\end{equation} 

Next, a comparison among various mass functions is presented. Instead of directly solving Eq. \eqref{ZEqnNum883658} which is challenging, we compare a transformed function $F_{X} \left(t\right)$ for given mass function $f_{X} (\nu)$, 
\begin{equation}
\label{ZEqnNum954771} 
F_{X} \left(t\right)=\int _{0}^{\infty }f_{X} \left(\nu \right) e^{-\left(\nu +v_{h} \right)t} d\nu ,         
\end{equation} 
where subscript $X$ is the abbreviation of the mass function model, i.e.  $X=PS,ST,D\lambda $. The function $\nu _{h} \left(m_{h} \right)$ in Eq. \eqref{ZEqnNum385449}) can be obtained using the fitting function in Eq. \eqref{ZEqnNum933840}. The transformed function $F_{X} \left(t\right)$ is computed by numerically integrating Eq. \eqref{ZEqnNum954771} for three analytical mass functions. The exact $F_{X}(t)$ in Eq. \eqref{ZEqnNum883658} for different $\gamma$ and $\alpha=4/3$ is also plotted for comparison

\begin{figure}
\includegraphics*[width=\columnwidth]{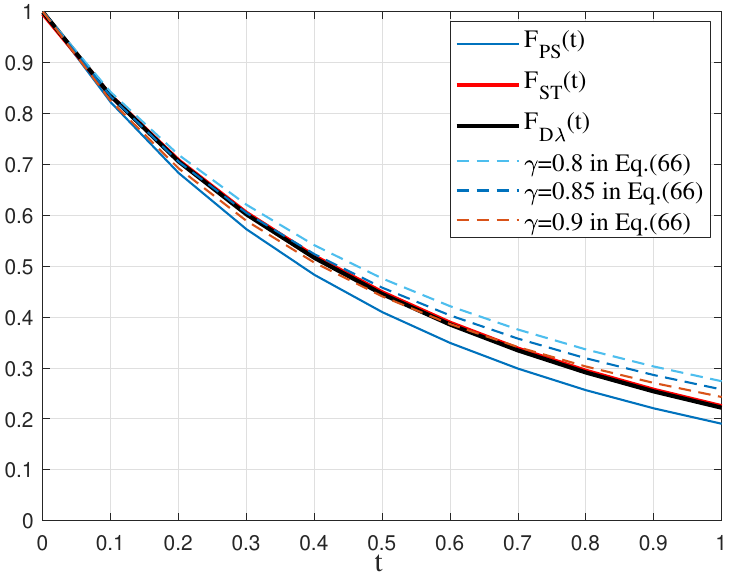}
\caption{The transformed function $F_{X} \left(t\right)$ for three different analytical mass functions (\textit{t} is a transformation variable and is not the physical time), compared against the analytical expression in Eq. \eqref{ZEqnNum883658} with $\alpha ={4/3} $, $v_{0}^{2} ={1/3} \sigma _{0}^{2} $, and $\gamma =0.8$, 0.85 and 0.9. The transformed functions $F_{D\lambda } \left(t\right)$ and $F_{ST} \left(t\right)$ almost coincide and match the target from Eq. \eqref{ZEqnNum883658} better than the PS mass function.}
\label{fig:4}
\end{figure}

Figure \ref{fig:4} plots the transformed function $F_{X} \left(t\right)$ for three different mass functions, compared against the analytical expression in Eq. \eqref{ZEqnNum883658} with $\alpha ={4/3} $, $v_{0}^{2} ={1/3} \sigma _{0}^{2} $, and $\gamma ={\alpha v_{0}^{2} /\bar{\sigma }_{h}^{2} } =0.8$, 0.85 and 0.9. The transformed functions $F_{D\lambda } \left(t\right)$ and $F_{ST} \left(t\right)$ almost coincide and agree better with the target $F_{X}(t)$ from Eq. \eqref{ZEqnNum883658} than the PS mass function. More study will be required for an exact solution of $f\left(\nu \right)$ from Eq. \eqref{ZEqnNum883658}, which will rely on an accurate model for halo velocity dispersion $\sigma _{h}^{2} $ or dimensionless dispersion $\nu _{h}$. 

\section{Conclusions}
\label{sec:4}
Halo mass function is a fundamental quantity for structure formation and evolution. Instead of basing mass functions on simplified spherical or elliptical collapse models, this paper attempts to interpret mass function as an intrinsic distribution to maximize system entropy of the everlasting statistically steady state in self-gravitating collisionless dark matter flow (SG-CFD). 

The limiting velocity (\textbf{\textit{X}}), speed (\textbf{\textit{Z}}), and energy (\textbf{\textit{E}}) distributions were previously obtained analytically from a maximum entropy principle \citep{Xu:2021-The-maximum-entropy-distributi}. In this paper, distributions of particle virial dispersion (\textbf{\textit{H}}), square of particle velocity (\textbf{\textit{P}}), and number of halos (\textbf{\textit{J}}) are proposed along with their connections with maximum entropy distribution (\textbf{\textit{X})}. The \textbf{\textit{H}} distribution for particle virial dispersion is essentially the halo mass function. By studying two limiting cases of \textbf{\textit{H}} distribution for large and small halos, we demonstrate that $H_{\infty } $ for large halos is also a maximum entropy distribution. For small halos, $H_{s}$ can be approximated by the distribution of square of particle velocity (\textbf{\textit{P}}). The $H_{s}$ distribution recovers the Press-Schechter mass function for small halos. The full solution of \textbf{\textit{H}} distribution depends on the limiting distribution (\textbf{\textit{X}}) that maximizes system entropy and the exact models for mass dependence of halo dispersion (Eqs. \eqref{ZEqnNum803629} and \eqref{ZEqnNum883658}). Future work includes better model of dispersion and more accurate solutions for \textbf{\textit{H}} distribution.


\section*{Data Availability}
Two datasets underlying this article, i.e. a halo-based and correlation-based statistics of dark matter flow, are available on Zenodo \citep{Xu:2022-Dark_matter-flow-dataset-part1,Xu:2022-Dark_matter-flow-dataset-part2}, along with the accompanying presentation slides "A comparative study of dark matter flow \& hydrodynamic turbulence and its applications" \citep{Xu:2022-Dark_matter-flow-and-hydrodynamic-turbulence-presentation}. All data files are also available on GitHub \citep{Xu:Dark_matter_flow_dataset_2022_all_files}.

\bibliographystyle{mnras}
\bibliography{Papers}

\appendix
\section{Statistical properties of \textbf{X} and \textbf{Z} distributions}
\begin{table*}
\caption{Statistical properties of $H_{\infty}$ and \textit{P} distributions}
\begin{tabular}{p{1.5in}p{2.2in}p{2.8in}} 
\hline 
Distribution Name & $H_{\infty }\left(x\right)$ & $P\left(x\right)$\\
\hline
Support & $[0,+\infty )$ & $[0,+\infty )$\\
\hline
PDF & $\frac{1}{2\alpha v_{0}^{2} K_{1} \left(a\right)} \cdot \exp \left[-\frac{\alpha }{2} \left(\frac{x}{\alpha v_{0}^{2} } +\frac{\alpha v_{0}^{2} }{x} \right)\right]$ & $\frac{e^{-\sqrt{\alpha ^{2} +{x/v_{0}^{2} } } } }{2\alpha v_{0} K_{1} \left(\alpha \right)\sqrt{x} } $ \\
\hline
CDF & $\frac{\alpha }{4K_{1} \left(\alpha \right)} \sum _{n=0}^{\infty }\frac{\left(-1\right)^{n} \alpha ^{2n} }{n!4^{n} }  \Gamma \left(-1-n,\frac{\alpha ^{2} v_{0}^{2} }{2x} \right)$ & $\frac{1}{\alpha K_{1} \left(\alpha \right)} \left[J_{\alpha } \left(\sqrt{\alpha ^{2} +{x/v_{0}^{2} } } \right)+\sqrt{\frac{x}{v_{0}^{2} } } \exp (-\sqrt{\alpha ^{2} +{x/v_{0}^{2} } } )\right]$\\
\hline
Mode & $\alpha v_{0}^{2} $ & 0 \\
\hline
Mean & $\frac{K_{2} \left(\alpha \right)}{K_{1} \left(\alpha \right)} \alpha v_{0}^{2} $ & $\frac{K_{2} \left(\alpha \right)}{K_{1} \left(\alpha \right)} \alpha v_{0}^{2} $\\
\hline
Variance & $\left[\frac{K_{3} \left(\alpha \right)}{K_{1} \left(\alpha \right)} -\frac{K_{2} \left(\alpha \right)^{2} }{K_{1} \left(\alpha \right)^{2} } \right]\left(\alpha v_{0}^{2} \right)^{2} $ & $\left[3\frac{K_{3} \left(\alpha \right)}{K_{1} \left(\alpha \right)} -\frac{K_{2} \left(\alpha \right)^{2} }{K_{1} \left(\alpha \right)^{2} } \right]\left(\alpha v_{0}^{2} \right)^{2} $\\
\hline
Moments & $\frac{K_{1+n} \left(\alpha \right)}{K_{1} \left(\alpha \right)} \left(\alpha v_{0}^{2} \right)^{n} $ & $\frac{K_{1+n} \left(\alpha \right)}{K_{1} \left(\alpha \right)} \frac{\Gamma \left(n+{1/2} \right)}{\sqrt{\pi } } \left(2\alpha v_{0}^{2} \right)^{n} $\\
\hline
Generalized kurtosis & $\frac{K_{1+n} \left(\alpha \right)}{K_{1} \left(\alpha \right)} \left(\frac{K_{1} \left(\alpha \right)}{K_{3} \left(\alpha \right)} \right)^{{n/2} } $ & $\frac{K_{1+n} \left(\alpha \right)}{K_{1} \left(\alpha \right)} \left(\frac{K_{1} \left(\alpha \right)}{K_{3} \left(\alpha \right)} \right)^{{n/2} } \frac{\Gamma \left(n+{1/2} \right)}{\sqrt{\pi } } \left(\frac{2}{\sqrt{3} } \right)^{n} $\\
\hline
Entropy & $-1+\alpha \frac{K_{2} \left(\alpha \right)}{K_{1} \left(\alpha \right)} +\ln \left(2\alpha v_{0}^{2} K_{1} \left(\alpha \right)\right)$ & $\ln \left(2\alpha v_{0}^{2} K_{1} \left(\alpha \right)\right)+\left(\alpha +\frac{1}{2\alpha } \right)\frac{K_{0} \left(\alpha \right)}{K_{1} \left(\alpha \right)} -\frac{1}{2} \left(\gamma +\ln 2-\ln \alpha \right)+1$\\
\hline
Moment-generating function & $\frac{K_{1} \left(\alpha \sqrt{1-2v_{0}^{2} t} \right)}{K_{1} \left(\alpha \right)\sqrt{1-2v_{0}^{2} t} } $ & $\sum _{n=0}^{\infty }\left(\begin{array}{c} {2n} \\ {n} \end{array}\right)\frac{K_{1+n} \left(\alpha \right)}{K_{1} \left(\alpha \right)} \left(\frac{\alpha v_{0}^{2} t}{2} \right)^{n}  $\\
\hline
Characteristic function & $\frac{K_{1} \left(\alpha \sqrt{1-2iv_{0}^{2} t} \right)}{K_{1} \left(\alpha \right)\sqrt{1-2iv_{0}^{2} t} } $ & $\sum _{n=0}^{\infty }\left(\begin{array}{c} {2n} \\ {n} \end{array}\right)\frac{K_{1+n} \left(\alpha \right)}{K_{1} \left(\alpha \right)} \left(i\frac{\alpha v_{0}^{2} t}{2} \right)^{n}  $\\
\hline
Maximum Entropy Constraints & $E\left(\frac{x}{2v_{0}^{2} } +\frac{\alpha ^{2} v_{0}^{2} }{2x} \right)=\alpha \frac{K_{2} \left(\alpha \right)}{K_{1} \left(\alpha \right)} -1$& \\
\hline

\end{tabular}
\label{tab:A1}

\begin{equation}
\begin{split}
&\textrm{Here } J_{s} \left(x\right) \textrm{ function is defined as the integral:}\\
&J_{s} \left(x\right)=\int _{s}^{x}e^{-t} \sqrt{t^{2} -s^{2} }  dt, \quad J_{s} \left(s\right)=0 \quad \textrm{and} \quad J_{s} \left(\infty \right)=sK_{1} \left(s\right)
\end{split}
\end{equation}

\begin{equation}
P\left(t\right)=\int _{0}^{\infty }X\left(x\right) e^{xt} dx=\frac{v_{0} te^{-\alpha } \left(1+\alpha \right)+J_{\alpha \sqrt{1-\left(v_{0} t\right)^{2} } } \left(\alpha \right)}{2\alpha K_{1} \left(\alpha \right)\left[1-\left(v_{0} t\right)^{2} \right]} +\frac{K_{1} \left(\alpha \sqrt{1-\left(v_{0} t\right)^{2} } \right)}{2K_{1} \left(\alpha \right)\sqrt{1-\left(v_{0} t\right)^{2} } }
\end{equation}

\begin{equation}
\begin{split}
&M_{Z} \left(t\right)=\int _{0}^{\infty }Y\left(x\right) e^{xt} dx=2\left[P\left(t\right)+t\frac{\partial P}{\partial t} \right]\\ 
&\textrm{Euler Constant } \gamma \approx 0.5772
\end{split}
\end{equation}

\begin{equation}
\begin{split}
&\textbf{\textit{J}} \textrm{ function is defined as the integral:}\\
&J\left(x\right)=\int _{1}^{x}e^{-t} \sqrt{t^{2} -1}  dt=\sum _{n=0}^{\infty }\left(-1\right)^{n}  \Gamma \left(\frac{3}{2} \right)\frac{\Gamma \left(2-2n,1\right)-\Gamma \left(2-2n,x\right)}{\Gamma \left(1+n\right)\Gamma \left({3/2} -n\right)} \\
&\textrm{where } \Gamma \left(x\right) \textrm{ is gamma function and}\\
&\Gamma \left(a,x\right) \textrm{ is an incomplete gamma function.}\\
&J\left(1\right)=0 \quad \textrm{and} \quad J\left(\infty \right)=K_{1} \left(1\right)
\end{split}
\end{equation}

\begin{equation}
\textrm{Constant} \quad \kappa =\int _{1}^{\infty }e^{-t} \sqrt{t^{2} -1} \ln \left(t\right) dt\approx 0.5333
\end{equation}

\end{table*}

\label{lastpage}
\end{document}